\documentclass[final,3p,times]{elsarticle}
\usepackage{hyperref}
\usepackage{amssymb}
\usepackage{amsmath}

\usepackage{float}
\usepackage{soul}
\usepackage[export]{adjustbox}
\usepackage{graphicx}
\usepackage{xcolor}
\usepackage[numbers]{natbib}  % For numbered references like [1]
\usepackage{caption}

\usepackage{color}

\usepackage{titlesec}
\titleformat{\paragraph}[block]{\normalfont\normalsize\itshape}{\theparagraph}{1em}{}

\usepackage[utf8]{inputenc}
\usepackage[T1]{fontenc}
\usepackage{textcomp}
\usepackage{lmodern}
\journal{arXiv } 
\begin{document}

\begin{frontmatter}

\title{A Comparative Review of Parallel Exact, Heuristic, Metaheuristic, and Hybrid Optimization Techniques for the Traveling Salesman Problem}

\author[iau]{Dr. Rabab Alkhalifa\corref{cor1}}
\cortext[cor1]{Corresponding author}
\ead{raalkhalifa@iau.edu.sa}

\author[iau]{Fatima Alkhomayes}
\author[iau]{Boushra Almazroua}
\author[iau]{Dana Alhaidan}
\author[iau]{Maryam Alothman}
\author[iau]{Jumana Almuhaidib}

% Boushra Almazroua 
% 2210003387@iau.edu.sa
% Fatima Alkhomayes 
% 2210002417@iau.edu.sa
% Jumana Almuhaidib 
% 2210002620@iau.edu.sa
% Maryam Alothman 
% 2210009016@iau.edu.sa
% Dana Alhaidan 
% 2210002781@iau.edu.sa 

\address[iau]{College of Computer Science and Information Technology, Imam Abdulrahman Bin Faisal University, P.O. Box 1982, Dammam 31441, Saudi Arabia}

\begin{abstract}
The Traveling Salesman Problem (TSP) is a well-known NP-hard combinatorial optimization problem with wide-ranging applications in logistics, routing, and intelligent systems. Due to its factorial complexity, solving large-scale instances requires scalable and efficient algorithmic frameworks, often enabled by parallel computing. This literature review provides a comparative evaluation of parallel TSP optimization methods, including exact algorithms, heuristic-based approaches, hybrid metaheuristics, and machine learning-enhanced models. In addition, we introduce task-specific evaluation metrics to facilitate cross-paradigm analysis, particularly for hybrid and adaptive solvers. The review concludes by identifying research gaps and outlining future directions, including deep learning integration, exploring quantum-inspired algorithms, and establishing reproducible evaluation frameworks to support scalable and adaptive TSP optimization in real-world scenarios.
\end{abstract}
%
% Our findings show that emerging machine learning-based solvers show promise, yet remain limited by generalization gaps, inference overhead, and the absence of standardized benchmarking protocols.

%% Keywords

\end{frontmatter}

%% sections
\section{Introduction}
\label{sec_intro}

The Traveling Salesman Problem (TSP) is one of the most well-known NP-hard problems in combinatorial optimization, with applications in logistics, urban planning, telecommunications, and bioinformatics. Formally, it is defined on a complete weighted graph \(G = (V, E)\), where each vertex \(v \in V\) represents a city, and each edge \(e_{ij} \in E\) has an associated cost \(C_{ij}\) representing the distance or travel expense between cities \(i\) and \(j\). The objective is to find a Hamiltonian cycle—a closed tour that visits every city exactly once—with minimal total travel cost:

\[
\min_{\pi} \sum_{i=1}^n C_{\pi_i, \pi_{i+1}}, \quad \text{where } \pi_{n+1} = \pi_1,
\]

and \(\pi\) is a permutation of the cities.

Due to its factorial growth (\(O(n!)\)), solving large-scale TSP instances using traditional methods like dynamic programming or branch-and-bound becomes computationally infeasible \cite{tsp_problem2006}. State-of-the-art solvers such as Concorde \cite{applegate2006concorde} can handle moderately sized problems but face exponential time complexity in the worst case.

\begin{figure}[htbp]
    \centering
    \includegraphics{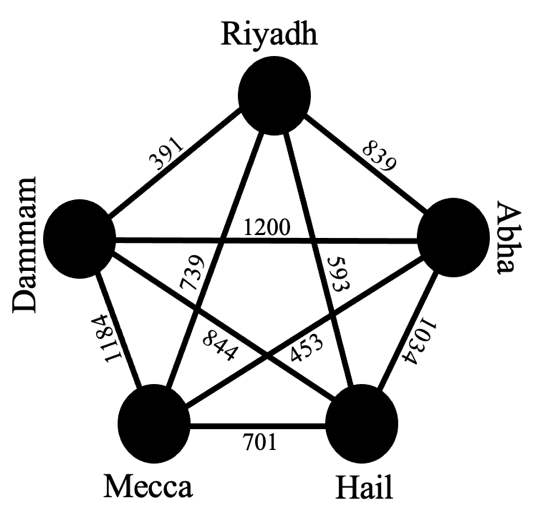}
    \caption{Modeling intercity distances as weighted edges in a complete graph. Example cities across Saudi Arabia serve as vertices with edges weighted by travel costs.}
    \label{fig1}
\end{figure}

Figure~\ref{fig1} illustrates the problem using Saudi cities, highlighting the combinatorial explosion as the number of cities increases. To overcome these computational challenges, parallel optimization strategies have been increasingly adopted.

The three main categories of parallel approaches for TSP includes:

\begin{enumerate}
    \item \textbf{Parallel Exact Algorithms:} Distribute exhaustive search (e.g., branch-and-bound) across processors for efficiency.
    \item \textbf{Parallel Heuristic and Metaheuristic Algorithms:} Approximation methods like genetic algorithms, ant colony optimization, and simulated annealing are parallelized to accelerate convergence.
    \item \textbf{Hybrid Metaheuristic Approaches:} Combine exact and heuristic techniques, often using heterogeneous hardware (CPUs/GPUs) to balance accuracy and performance.
\end{enumerate}

Each method must manage trade-offs in synchronization, memory access, and communication overhead. Advances in multi-core processors, GPUs, distributed clusters, and cloud platforms have enabled significant reductions in computation time for large TSP instances \cite{dahiya2018literature}.

\vspace{0.5em}
\noindent \textbf{Research Questions}

Despite substantial progress, existing research lacks consistent benchmarks and performance metrics across parallel TSP solvers. This review addresses that gap by investigating state-of-the-art approaches in parallel optimization for TSP, structured around the following research questions:

\begin{itemize}
% \item[\textbf{RQ1}] What is the formal definition of the Traveling Salesman Problem, and how is parallelism conceptualized and applied to TSP optimization? \textbf{Section~\ref{sec_intro}}
\item[\textbf{RQ1}] What are the main categories of parallel algorithms used for solving TSP—including exact, heuristic, metaheuristic, and hybrid approaches—and what are the underlying methodologies of each? \textbf{Section~\ref{sec_approch}}
\item[\textbf{RQ2}] What datasets are commonly employed to evaluate parallel TSP solvers, and how do dataset characteristics (size, topology, cost metrics) influence algorithm performance?  \textbf{Section~\ref{sec_datasets}}
\item[\textbf{RQ3}] What evaluation metrics are used to assess the performance of parallel TSP optimization methods across different algorithmic categories, particularly regarding execution speed, solution quality, scalability, and robustness? \textbf{Section~\ref{sec_eval}}
\item[\textbf{RQ4}] How are emerging computational paradigms, such as machine learning (including deep learning) and quantum computing, being integrated into parallel TSP optimization, and what future trends do they suggest for enhancing scalability and efficiency? \textbf{Section~\ref{sec_gap}}
\end{itemize}

\vspace{0.5em}

The remainder of this paper is structured as follows: \textbf{Section~\ref{sec_approch}} presents the various parallel optimization techniques, including exact methods, heuristic-based approaches, and hybrid metaheuristics (\textbf{RQ1}). \textbf{Section~\ref{sec_datasets}} reviews commonly used benchmark datasets for evaluating parallel TSP solvers, addressing dataset characteristics and their impact on algorithm performance (\textbf{RQ2}). \textbf{Section~\ref{sec_eval}} introduces evaluation metrics for comparing solver performance across execution speed, scalability, accuracy, and memory usage (\textbf{RQ3}). \textbf{Section~\ref{sec_gap}} identifies research gaps and explores future directions, including the integration of AI-based heuristics and quantum computing techniques (\textbf{RQ4}). Finally, \textbf{Section~\ref{sec_conc}} concludes the review and summarizes key findings and recommendations.

% The remainder of this paper is structured as follows: \textbf{Section~\ref{sec_approch}} presents the various parallel optimization techniques, including exact methods, heuristic-based approaches, and hybrid metaheuristics (\textbf{RQ1}). \textbf{Section~\ref{sec_datasets}} reviews commonly used benchmark datasets for evaluating parallel TSP solvers, addressing dataset characteristics and their impact on algorithm performance (\textbf{RQ2}). \textbf{Section~\ref{sec_eval}} introduces evaluation metrics for comparing solver performance across execution speed, scalability, accuracy, and memory usage (\textbf{RQ3}). \textbf{Section~\ref{sec_gap}} identifies research gaps and explores future directions, including the integration of AI-based heuristics and quantum computing techniques (\textbf{RQ4}). Finally, \textbf{Section~\ref{sec_conc}} concludes the review and summarizes key findings and recommendations.

% \input{sections/sec2}
\section{Parallel Optimization Techniques in TSP}
\label{sec_approch}
The TSP is a classical combinatorial optimization problem in which the objective is to find the shortest possible route that visits each city in a given set exactly once and returns to the starting point. Due to its factorial growth in solution space ($O(n!)$), the TSP becomes computationally intractable as the number of cities increases, posing a significant challenge for large-scale instances.

This section focuses on three primary categories of parallel optimization techniques developed to address the computational demands of the TSP:

\begin{enumerate}
    \item \textbf{Parallel Exact Algorithms} — Guarantee optimal solutions but incur exponential time and memory complexity.
    \item \textbf{Parallel Heuristic and Metaheuristic Algorithms} — Provide high-quality approximate solutions efficiently using methods such as Genetic Algorithms (GA) and Ant Colony Optimization (ACO).
    \item \textbf{Hybrid Metaheuristic Approaches} — Combine multiple optimization paradigms (e.g., heuristics, metaheuristics, and learning-based strategies) to improve scalability, convergence speed, and robustness.
\end{enumerate}

\subsection{\textbf{Parallel Exact Algorithms}}
\label{sec3.1}
% \todo{https://www.sciencedirect.com/science/article/pii/S187705091732375X}

Exact TSP solvers systematically explore the solution space to guarantee finding the optimal tour. However, these methods are severely constrained by their exponential time complexity, which renders them impractical for large-scale instances. Even advanced approaches like dynamic programming \cite{alquaamiz2024parallelization} and integer linear programming (ILP) \cite{dell2020matheuristic} face prohibitive memory and computational costs.

Parallelization of exact methods provides a natural way to alleviate these limitations by distributing the computational burden across multiple processing units. One widely used method is the branch-and-bound (B\&B) algorithm, where each node of the search tree corresponds to a partial tour. This structure lends itself well to parallel exploration \cite{burkhovetskiy2017parallelizing}.

Three major strategies have emerged for parallelizing exact TSP solvers:

\begin{itemize}
\item \textbf{Task Decomposition:} The search space is divided into independent subproblems, each assigned to a different processor or thread. Each unit explores its subtree using the same B\&B logic, with only periodic coordination needed to share the global best solution (upper bound) \cite{tschoke1995solving}. For example, Burkhovetskiy et al. \cite{burkhovetskiy2017parallelizing} adapted the Balas and Christofides algorithm for parallel processing, achieving a 2.08x speedup on instances with up to 2,000 cities. While effective, their results also revealed diminishing returns due to synchronization overhead and load imbalance

\item \textbf{Shared-Memory Architectures:} On multi-core CPUs, threads typically share a global work pool and a common record of the best solution found. Load balancing is achieved via dynamic scheduling or work-stealing strategies. For instance, parallelized B\&B using OpenMP, significantly reducing computation time on large TSP instances \cite{al2023parallel}. However, shared-memory models demand thread-safe data structures and can suffer from contention in accessing shared memory, especially when updates to the global best cost require locking mechanisms. Performance also degrades due to cache thrashing when multiple threads access disparate memory locations. Designers often mitigate this by assigning spatially close subproblems to threads, improving data locality and reducing cache misses \cite{barai2019modeling}.

\item \textbf{Distributed-Memory Architectures:} In cluster or cloud environments without shared memory, processors must communicate via message passing (e.g., MPI). Each node solves a portion of the search tree and periodically shares its best-found solutions to aid global pruning. Strategies like \textit{Parallel Synchronized Branch-and-Bound (PSBB)} \cite{Diderich1996} synchronize only during key phases to minimize overhead. Other designs rely on \textit{distributed global priority queues} \cite{clausen1999branch} to simulate best-first traversal across nodes. Load balancing is a persistent challenge: idle nodes must dynamically acquire tasks from busy ones, requiring sophisticated communication protocols. 

Additionally, branch-and-cut algorithms—which extend B\&B by dynamically adding cutting planes to tighten LP relaxations—can exploit distributed architectures. Subtasks such as constraint generation and LP solving can be performed concurrently \cite{Ladanyi2001}. Still, these routines are often memory-bound, and their parallel scalability plateaus with increasing node counts due to solver-level synchronization \cite{maher2021assessing}.

\end{itemize}

In both shared- and distributed-memory environments, the overarching challenge lies in maintaining high parallel efficiency. This involves minimizing idle time, optimizing memory access patterns, and reducing the cost of synchronization and communication. Studies such as \cite{alquaamiz2024parallelization,gohil2022travelling} report near-linear speedups for carefully tuned parallel B\&B implementations, provided that strong initial upper bounds and adaptive load balancing are used.

Despite ongoing efforts to parallelize exact methods, their scalability remains limited. This limitation has driven growing interest in heuristic and metaheuristic approaches, which offer scalable and efficient alternatives for obtaining near-optimal solutions. While exact solvers remain valuable for small to medium instances or when optimality is critical, their heavy resource requirements underscore the need for more scalable optimization paradigms \cite{rokbani2021bi}.

\subsection{\textbf{Parallel Heuristic and Metaheuristic Algorithms}}
\label{sec3.2}

Heuristic and metaheuristic algorithms constitute two important classes of approximate solution methods widely employed for the TSP, particularly when exact algorithms become computationally prohibitive for large instances. While often used interchangeably, they differ fundamentally in scope, design, and generality:

\begin{itemize}

\item \textbf{Heuristics} are problem-specific algorithms or rules designed to quickly construct or improve a single solution using domain knowledge or simple local search principles. They typically operate with polynomial time complexity and exploit specific structural properties of the problem to find reasonably good solutions efficiently. Examples include nearest neighbor heuristics, insertion heuristics, and local search techniques such as 2-opt and 3-opt. Heuristics focus on intensification — efficiently exploiting the local neighborhood of a current solution to improve it, often making greedy or deterministic decisions. 

\item \textbf{Metaheuristics} provide higher-level, problem-independent frameworks that guide heuristic or stochastic search processes in a strategic manner. These algorithms operate over populations or multiple candidate solutions and emphasize a balance between exploration (searching broadly across the solution space) and exploitation (refining promising solutions). Metaheuristics incorporate mechanisms such as probabilistic transitions, evolutionary operators, memory structures, or pheromone communication to avoid premature convergence to local optima. Examples include Genetic Algorithms, Ant Colony Optimization, Simulated Annealing, and Tabu Search.

\end{itemize}

The primary advantage of metaheuristics is their generality: they can be adapted to a wide range of combinatorial optimization problems with minimal problem-specific tuning. Moreover, metaheuristics typically operate iteratively or population-wise, which naturally lends itself to parallelization—multiple candidate solutions can be evaluated or evolved simultaneously. In the context of parallel TSP optimization, these distinctions manifest as follows:

\begin{itemize}
\item \textbf{Heuristic algorithms}, being often sequential and constructive or local in nature, parallelize primarily by concurrently improving or constructing multiple independent solutions or by parallelizing computations within the local search neighborhood evaluation. For example, a parallel 2-opt heuristic can simultaneously evaluate many possible edge swaps across different segments of the tour to benefit from data-parallel evaluations but suffer from dependency and convergence synchronization.

\item \textbf{Metaheuristic algorithms} inherently support coarse-grained parallelism by evolving populations (GA) or colonies of agents (ACO) concurrently. Parallelism is realized by distributing individuals or agents to different processors or threads, enabling simultaneous solution construction, evaluation, and genetic or pheromone updates.
\end{itemize}

\subsubsection{\textbf{Genetic Algorithms (GA)}}
\label{sec3.2.1}
Genetic Algorithms are population-based evolutionary techniques inspired by natural selection and genetics. They iteratively improve candidate solutions through selection, crossover, and mutation operations, balancing exploration and exploitation of the search space. Parallel GA implementations typically employ the following strategies to enhance computational efficiency:

\begin{itemize}

\item \textbf{Island Model GA}: The population is partitioned into subpopulations (islands) that evolve independently on separate processors, with occasional migration of individuals between islands to maintain genetic diversity and avoid premature convergence \cite{da2019parallel}.

\item \textbf{Master-Slave GA}: A master node orchestrates genetic operations, delegating fitness evaluations of candidate solutions to multiple worker nodes that operate in parallel \cite{peng2022parallel}.

\end{itemize}

GPU-accelerated GA implementations using CUDA and OpenMP have achieved up to 60\% faster execution than CPU-only versions, maintaining approximately 98\% near-optimal solution quality \cite{abbasi2020efficient}. These gains position GA as one of the most effective heuristics for large-scale TSP problems.

\subsubsection{\textbf{Ant Colony Optimization (ACO)}}
\label{sec3.2.2}
ACO is a swarm intelligence metaheuristic inspired by the pheromone-based foraging behavior of ants. Artificial ants construct solutions probabilistically by following and reinforcing pheromone trails, which guide subsequent ants toward better routes \cite{katiyar2015ant}. This method adapts dynamically based on historical success and is particularly suited for dynamic and uncertain optimization environments. Parallel ACO implementations commonly include:

\begin{itemize}
\item \textbf{Multi-Colony ACO}: Multiple ant colonies explore solution spaces in parallel and periodically exchange information to enhance diversity and prevent premature convergence \cite{delevacq2013parallel}.
\item \textbf{GPU-Optimized ACO}: Using CUDA and OpenMP, these implementations parallelize both solution construction and pheromone update phases, achieving speedups of up to 43× compared to CPU-based versions \cite{yang2024tensorized}.
\end{itemize}

ACO’s tendency toward premature convergence from excessive pheromone accumulation is addressed through adaptive evaporation and parameter tuning strategies, maintaining efficient exploration \cite{mavrovouniotis2013evolving}.

\subsubsection{\textbf{Simulated Annealing (SA)}}
\label{sec3.2.3}
Simulated annealing is a probabilistic optimization algorithm inspired by the physical annealing process, where controlled cooling allows materials to settle into low-energy states. SA iteratively accepts new solutions based on a temperature-dependent probability, enabling it to escape local optima and move toward near-global optima \cite{kirkpatrick1983optimization}. Its robustness makes it effective for both combinatorial and continuous optimization problems. Parallel SA techniques include:

\begin{itemize}
\item \textbf{Multi-Start SA}: Multiple independent SA instances run concurrently from different initial conditions, increasing the likelihood of finding the global optimum and reducing runtime \cite{sonuc2018cooperative}.
\item \textbf{Hybrid Parallel SA}: SA is combined with evolutionary algorithms or local search methods to enhance convergence speed and solution quality \cite{borisenko2017parallelizing}.
\item \textbf{GPU-Accelerated SA}: CUDA and OpenCL implementations speed up computationally intensive steps such as solution evaluation and acceptance testing, improving performance on large-scale problems \cite{de2018experimental}.
\end{itemize}

The cooling schedule critically influences SA’s performance; adaptive cooling and self-tuning mechanisms are employed to balance diversification and intensification for effective convergence \cite{sereapplication}.

\subsubsection{\textbf{Tabu Search (TS) }}
\label{sec3.2.4}

Tabu Search (TS) is a powerful metaheuristic designed to escape local optima by using a short-term memory structure called the \emph{tabu list}, which forbids or penalizes recently visited solutions to prevent cycling. This allows TS to explore the solution space more thoroughly by balancing diversification and intensification \cite{glover1989tabu}.

Parallel TS methods can be broadly categorized into two main types, enhancing performance on large-scale problems like TSP:

\begin{itemize}
    \item \textbf{Parallel Multi-Start TS}: Multiple Tabu Search instances are executed in parallel, each initialized with a different starting solution. These independent searches may periodically exchange elite solutions—typically the best-so-far tours—to enhance global exploration and prevent premature convergence. Depending on the architecture, these instances may run as threads, processes, or distributed agents. Information exchange can be synchronous (at fixed intervals) or asynchronous (on-demand), with asynchronous models offering better scalability. Although less common, partial sharing of tabu memory structures has also been explored but must be used cautiously to preserve search independence and avoid search bias \cite{talbi2009metaheuristics, talbi2002taxonomy}.

    \item \textbf{Parallel Neighborhood Evaluation TS}: This approach accelerates Tabu Search by parallelizing the evaluation of candidate moves within each neighborhood iteration. It employs data parallelism by dividing the set of possible moves among multiple processors or threads, allowing simultaneous computation of move costs or fitness values. This is particularly beneficial for large-scale TSP instances where neighborhood sizes grow rapidly. After local evaluations, a synchronization step is typically required to identify the globally best admissible move, which may introduce communication overhead. More advanced implementations leverage speculative evaluation or GPU acceleration to mitigate such bottlenecks \cite{hou2020efficient,Pardalos2002parallel}.

\end{itemize}

Parallel TS has shown significant improvements in solution quality and runtime for large problems, achieving near-linear scaling up to dozens of cores \cite{crainic2005parallel}. However, challenges such as tabu list management, communication overhead, and ensuring diversity in the search space remain \cite{DudekDyducb1995}.

In summary, heuristic and metaheuristic approaches offer scalable and efficient solutions for large-scale TSPs where exact methods are impractical. Parallel computing advancements, such as GPU acceleration and multi-core processing, have further accelerated these algorithms and improved solution quality. However, challenges like maintaining solution diversity and avoiding premature convergence persist. Future work is expected to explore hybrid heuristics combining these methods to further enhance optimization across diverse domains.

\subsection{\textbf{Hybrid Metaheuristics \& Emerging Techniques}}
\label{sec3.3}
Hybrid metaheuristic approaches combine multiple optimization algorithms to enhance solution quality, convergence speed, and computational efficiency. These hybrid techniques leverage the strengths of individual heuristics while mitigating their limitations, making them highly effective for solving large-scale TSP instances. 

The recent integration of evolutionary computing, swarm intelligence, and AI-driven strategies has substantially advanced the state-of-the-art in TSP optimization. The following summarizes some of the most promising hybridization techniques, supported by representative studies:

\begin{itemize}
    \item \textbf{GA-ACO Hybridization}: This approach combines GA for global exploration with ACO for local exploitation, balancing diversification and intensification more effectively than either alone. GA produces diverse candidate solutions, while ACO refines these using pheromone-guided search to accelerate convergence and enhance solution quality. \cite{soylu2017hybrid} demonstrated that this hybrid approach improves convergence speed and solution quality on benchmark TSP instances compared to standalone GA or ACO. Likewise,  \cite{gharehchopogh2012new} reported that their GA-ACO hybrid outperformed pure algorithms in both accuracy and computational efficiency on dynamic TSP problems.

    \item \textbf{ML Perspective on AI-Augmented Heuristics}: These approaches frame combinatorial optimization problems like the TSP as learning tasks rather than purely algorithmic search challenges. Instead of manually tuning heuristic parameters or relying on fixed search strategies, machine learning models—especially deep neural networks and reinforcement learning agents—learn policies that guide the optimization process dynamically. By framing TSP and related problems as Markov Decision Processes or sequence generation tasks, these ML-based methods shift the solution process from hand-crafted heuristics to learned policies. This allows models to generalize across instances, adapt to complex landscapes, and improve performance through experience, bridging the gap between data-driven learning and classical optimization.

    Early work by Vinyals et al. \cite{vinyals2015pointer} introduced Pointer Networks, leveraging attention mechanisms to model routing as a sequence prediction task. Building on this foundation, Kool et al. \cite{koolattention} proposed an attention-based deep reinforcement learning framework that constructs tours sequentially, learning effective heuristics directly from data. Parallel approaches utilize graph neural networks to capture the structural properties of routing problems more explicitly, as demonstrated by Joshi et al. \cite{joshi2019efficient}. Hybrid methods combining deep learning with classical heuristics have also shown promise; for instance, Xin et al. \cite{xin2021neurolkh} integrate learned models with the Lin-Kernighan-Helsgaun heuristic to enhance solution quality and efficiency. A recent survey by Yang and Whinston \cite{yang2023survey} highlights the growing trend of reinforcement learning methods in combinatorial optimization, emphasizing adaptive and data-driven heuristics as a key direction for future research.
    
    \item \textbf{Quantum-Inspired and Quantum Computing TSP Solvers}—Quantum computing introduces new ways to solve complex optimization problems, including TSP. One key technique, called quantum annealing, uses principles of quantum physics—such as superposition, where many possibilities exist simultaneously, and tunneling, which helps escape poor solutions—to explore multiple solution paths at once. This parallelism can potentially speed up the search for the best route compared to classical methods \cite{farhi2000quantum}.
    
    Building on these ideas, researchers have developed quantum-inspired algorithms that mimic quantum behaviors but run on regular computers. For example, quantum-inspired genetic algorithms represent solutions using quantum bits (qubits) and apply special quantum-inspired operators to maintain diverse and high-quality candidate solutions, improving how quickly and effectively the search converges \cite{Hyun2002QuantumInspired}.
    
    Physical quantum devices, like D-Wave’s quantum annealers, have been used experimentally to tackle small to medium TSP problems, showing promising improvements in solving time compared to traditional algorithms \cite{venturelli2015quantum, king2019quantum}. However, current quantum hardware still faces challenges such as limited size and sensitivity to noise, which restrict their ability to handle larger, real-world problems.

To overcome these limits, hybrid quantum-classical algorithms combine quantum computing’s unique capabilities with proven classical metaheuristics. These hybrid approaches allow leveraging the strengths of both worlds—using quantum devices for parts of the problem while classical algorithms handle others—offering a practical path toward more powerful and scalable optimization solvers.

\end{itemize}

\begin{table}[htbp]
\captionsetup{justification=raggedright, singlelinecheck=false}
\centering
\caption{Comparative Overview of Parallel Optimization Strategies for TSP}
\begin{adjustbox}{max width=0.99\textwidth}
\begin{tabular}{|p{3.2cm}|p{2cm}|p{2.2cm}|p{3.2cm}|p{4cm}|p{3.8cm}|}
\hline
\textbf{Category} & \textbf{Optimization Quality} & \textbf{Scalability} & \textbf{Aim} & \textbf{Strategy} & \textbf{Challenges} \\
\hline
 \textbf{Exact Algorithms} & Optimal & Low--Moderate & Minimize exhaustive search time through workload distribution & Search tree decomposition (e.g., B\&B, B\&C), ILP, DP & High memory usage, synchronization overhead, diminishing returns at scale \\
\hline
\textbf{Heuristic \& Metaheuristic Algorithms} & Near-optimal & High & Speed up solution construction and exploration processes & Local and global parallelism (e.g., parallel 2-opt, GA populations, ACO agents) & Premature convergence, parameter sensitivity, inter-process communication \\
\hline
\textbf{Hybrid \& Emerging Techniques} & High--Very High & High (scalable with future hardware) & Integrate complementary strategies for improved convergence and generalization & Fusion of diverse methods (e.g., GA+ACO), ML-guided policies, quantum-classical synergy & System complexity, integration overhead, algorithm-hardware mismatch \\
\hline
\end{tabular}
\end{adjustbox}
\label{tab:parallel_tsp_comparison}
\end{table}

A comparative overview of the main categories of parallel TSP optimization strategies is presented in Table~\ref{tab:parallel_tsp_comparison}. It highlights their respective optimization capabilities, scalability, parallel objectives, implementation approaches, and associated challenges. Among these, hybrid metaheuristics and emerging techniques represent a particularly promising research frontier. By integrating population-based heuristics (e.g., GA, ACO), data-driven learning (e.g., reinforcement learning, graph neural networks), and experimental quantum computing paradigms, researchers can develop solvers that better balance exploration and exploitation, dynamically adapt to problem structures, and leverage modern computational architectures.

% A concise comparison of these parallel optimization strategies is presented in Table~\ref{tab:parallel_tsp_comparison}, summarizing their optimization quality, scalability, parallel aims, and key implementation challenges. Overall, hybrid metaheuristics represent a vital research frontier for scalable and effective TSP solving. By integrating GA, ACO, ML-driven adaptivity, and emerging quantum methods, researchers can develop solvers that balance exploration-exploitation trade-offs, adapt dynamically to problem features, and leverage computational advances. Future work should explore hybrid parallelization strategies, deep reinforcement learning enhancements, and practical quantum-classical frameworks to further improve large-scale TSP optimization. 

% These directions align well with stepwise classification paradigms, which model TSP as sequential decision-making, suggesting promising opportunities for synergy between learned policies and metaheuristic refinement \todo{CIRE MY Work}.

\section{Benchmark Datasets for Parallel TSP Optimization}
\label{sec_datasets}
Standardized benchmark datasets are essential for ensuring reproducibility and comparability in TSP research. To fairly and consistently evaluate parallel TSP solvers—spanning exact, heuristic, metaheuristic, AI-driven, and quantum-inspired methods—researchers rely on well-established benchmark collections. These datasets comprise both theoretical and real-world instances, enabling comprehensive assessment of algorithmic efficiency, scalability, and robustness.

% As summarized in Table\ref{tab1}, key datasets commonly used in TSP research, along with their features and download references. While TSPLIB \cite{reinelt1991tsplib} serves as the fundamental benchmark for evaluating traditional and hybrid TSP solvers, the Mona Lisa dataset \cite{bosch2004continuous} challenges scalability for large-scale optimization techniques. Meanwhile, randomly generated TSP \cite{basel2001random} instances assess generalization capability, and national road network datasets ensure real-world applicability \cite{us2014usgs}.

\begin{table}[H]
\captionsetup{justification=raggedright, singlelinecheck=false}
\caption{Commonly Used TSP Datasets}
\label{tab1}
\centering
\renewcommand{\arraystretch}{1.2}
\begin{tabular}{|p{1cm}|p{3cm}|p{2.5cm}|p{2.5cm}|p{5cm}|}
\hline
\textbf{Ref.} & \textbf{Dataset} & \textbf{Number of Cities} & \textbf{Type} & \textbf{Applications} \\
\hline
\cite{tsplib} & \textbf{TSPLIB} & 14 -- 85{,}900 & Synthetic \& Real World & Validating exact, heuristic, and hybrid solvers. \\
\hline
\cite{monalisaTSP} & \textbf{Mona Lisa TSP} & 100{,}000 & Synthetic & Testing scalability and parallel efficiency of algorithms. \\
\hline
\cite{tspgen} & \textbf{Random TSP Instances} & User defined & Synthetic & Evaluating generalization and performance in varied scenarios. \\
\hline
\cite{usgsRoadNetwork} & \textbf{National Road Network Datasets} & Varies & Real World & Real-time route planning, logistics, and smart city applications. \\
\hline
\end{tabular}
\end{table}

As summarized in Table~\ref{tab1}, key datasets commonly used in TSP research include:

\begin{itemize}
    \item \textbf{TSPLIB} \cite{reinelt1991tsplib}, the foundational benchmark repository, provides a broad range of problem instances from small to large scales with realistic city coordinates and distance matrices. It remains the primary standard for evaluating traditional, hybrid, and parallel solvers.
    
    \item The \textbf{Mona Lisa dataset} \cite{bosch2004continuous} features over 10,000 nodes derived from pixel data of the Mona Lisa image. This large-scale, geometrically complex instance tests the scalability and efficiency of parallel and approximate algorithms.
    
    \item \textbf{Randomly generated TSP instances} \cite{basel2001random}, created under uniform or clustered spatial distributions, assess solvers’ generalization ability and robustness to varying problem structures.
    
    \item \textbf{National road network datasets} \cite{us2014usgs} capture real-world routing challenges by modeling large-scale transportation networks with realistic connectivity and asymmetric travel costs, providing critical evaluation grounds for practical logistics and route planning applications.
\end{itemize}

These datasets, collectively, offer a holistic evaluation framework that enables researchers to:

\begin{itemize}
    \item Develop, refine, and benchmark parallel TSP-solving techniques across diverse problem scales and complexities.
    \item Ensure fair, reproducible comparisons among algorithms from classical heuristics to cutting-edge AI-augmented and quantum-inspired methods.
    \item Drive progress in scalable, high-performance combinatorial optimization for both theoretical and applied contexts.
\end{itemize}

While TSPLIB and complementary datasets span a wide variety of problem types, their suitability depends on the solver approach. Exact methods are typically evaluated on small to medium TSPLIB instances, as they offer rigorous benchmarking and manageable complexity for optimal solvers such as Concorde. In contrast, heuristic and metaheuristic methods are often tested on large-scale instances, including the Mona Lisa dataset \cite{bosch2004continuous} and national road networks, to demonstrate scalability and robustness. AI-driven models—particularly those using neural networks—frequently rely on synthetic, randomly generated datasets for training and generalization assessment \cite{koolattention, xin2021neurolkh}.

As quantum computing approaches continue to develop, there is an emerging need for tailored small-to-medium scale benchmark instances that align with the hardware constraints of current quantum annealers and simulators.

In conclusion, the continuous expansion and diversification of benchmark datasets will remain crucial for advancing TSP optimization methodologies. There is growing interest in integrating AI-augmented heuristics and quantum-inspired algorithms, which will likely require new benchmarking standards to support consistent and meaningful evaluation of these hybrid approaches.

\section{Evaluation Measures for Parallel TSP Optimization}
\label{sec_eval}
The evaluation of TSP optimization techniques is inherently method-dependent since different algorithms operate under distinct computational paradigms.

\begin{itemize}
    \item \textbf{Exact algorithms} prioritize solution optimality but suffer from high computational complexity. Parallelism is limited by data dependencies, resulting in sublinear speedup. Large memory consumption and synchronization overhead reduce parallel efficiency.
    \item \textbf{Heuristic methods} prioritize computational efficiency but do not guarantee optimality. Parallelization is more effective as heuristics operate on independent solution candidates (e.g., GA population, ACO agents). However, communication overhead in pheromone updates can reduce efficiency.
    \item \textbf{Hybrid metaheuristic approaches}  integrate multiple techniques to balance accuracy and efficiency, often leveraging parallel computing for scalability. Hybrid methods can leverage parallelism efficiently by combining parallelizable heuristics with controlled exact computations. CUDA-accelerated hybrids achieve high parallel efficiency.
\end{itemize}

Thus, evaluation must account for the unique trade-offs associated with each approach, rather than relying on a single universal metric. This section introduces general performance indicators applicable across all methods and proposes method-specific evaluation metrics to highlight the strengths and limitations of exact, heuristic, and hybrid strategies. Such a classification provides a structured framework for comparative TSP solver assessment.

\subsection{\textbf{General Performance Metrics}}
\label{sec5.1}
These fundamental metrics apply to \textbf {exact algorithms}, \textbf {heuristics solvers}, and \textbf {hybrid metaheuristic approaches}, providing a basis for comparison.

\subsubsection{\textbf{Execution Time and Speedup}}
\label{sec5.1.1}

Execution time $T(n, p)$ is a fundamental measure of algorithmic efficiency, representing the total computation time required to solve a TSP instance of size $n$ using $p$ processors. It is critical for evaluating solver performance, particularly in parallel optimization settings.

Execution time strongly differentiates exact and heuristic approaches. \textbf{Exact algorithms} (e.g., branch-and-bound, dynamic programming) guarantee optimal solutions but exhibit exponential growth in execution time, making them infeasible for large $n$. For example, the Held-Karp dynamic programming algorithm has a time complexity of:

\begin{equation}
T(n) = \mathcal{O}(2^n \cdot n^2)
\label{eq1}
\end{equation}

\textbf{Heuristic solvers} trade optimality for significantly faster solutions by operating under lower complexity. These methods enable near-optimal solutions in practical time for large-scale problems. These algorithms respond differently to parallelization, with speedup and efficiency largely dependent on algorithm structure:

\begin{itemize}
    \item \textbf{GA} is highly parallelizable, particularly in evaluating the fitness of individuals within a population. GPUs can accelerate these evaluations, yielding substantial speedups. However, the actual gains depend on the complexity of the fitness function and the overhead introduced by kernel launches and memory access patterns \cite{zhuo2023parallel}.

    \item \textbf{ACO} supports parallel tour construction by agents (ants), but global pheromone update synchronization introduces communication overhead. To mitigate this, techniques such as partitioned search spaces and asynchronous updates have been used to enhance scalability without sacrificing accuracy \cite{janson2005parallel}.

    \item \textbf{SA} is inherently sequential because each new state depends on the current solution. Nevertheless, parallelism can be achieved by running multiple independent annealing chains or using speculative computation \cite{sohn1996generalized,witte1991parallel}. \textit{Generalized Speculative Computation (GSC)} has been reported to achieve over 20$\times$ speedup in domain-specific cases, such as the TSP, by enabling parallel execution while preserving the serial decision logic of SA~\cite{sohn1996generalized}

\end{itemize}

% witte1991parallel

% \paragraph{Hybrid Metaheuristics.}  
\textbf{Hybrid solvers} combine heuristic and local search techniques (e.g., GA with 2-opt or ACO with Tabu Search) to balance quality and efficiency. These hybrids often parallelize well at the component level. For instance, population-based search and local refinements can be computed in parallel, though coordination between components may introduce synchronization costs. 

% \hl{CUDA-accelerated hybrids have demonstrated runtime reductions in fitness evaluation and local search modules} \hl{[needs stronger citation]}.

% \paragraph{Speedup and Practical Challenges.}  
In parallel computing, speedup $S$ is defined as:

\begin{equation}
S = \frac{T(n, 1)}{T(n, p)}
\label{eq2}
\end{equation}

\noindent where $T(n, 1)$ and $T(n, p)$ denote the execution time using one and $p$ processors, respectively. Ideal linear speedup ($S = p$) is rarely achieved due to synchronization delays, data dependencies, and memory contention \cite{yavits2014effect, bertuletti2023fast, giannoula2021syncron}.

% \paragraph{Evaluation Concerns.}  
While execution time is a critical metric, it is influenced by multiple factors—problem size, hardware architecture, memory bandwidth, and implementation quality. This makes direct comparison across solvers challenging. Moreover, maximizing speedup often requires careful algorithm design to ensure load balancing, minimize inter-process communication, and fully utilize hardware parallelism.

\subsubsection{\textbf{Scalability Metrics}}
\label{sec5.1.2}

Scalability refers to an algorithm’s ability to maintain efficiency as problem size or processor count increases. It is a critical factor in the evaluation of parallel TSP solvers, particularly when applied to large-scale optimization tasks. Some algorithms perform well on small instances but experience performance degradation as the problem grows, often due to communication overhead, memory bottlenecks, or task imbalance.

Two standard scalability metrics are commonly used:

\begin{itemize}
    \item \textbf{Strong Scaling Efficiency:}
        \begin{equation}
        E_s = \frac{T(n, 1)}{P \cdot T(n, P)}
        \label{eq3}
        \end{equation}
        This measures how well an algorithm speeds up when the number of processors increases for a fixed problem size. Ideally, $E_s \rightarrow 1$ as $P$ increases. In practice, strong scaling efficiency decreases due to communication costs, synchronization overhead, and reduced computation per processor. This effect is particularly pronounced in exact solvers with tight data dependencies.

    \item \textbf{Weak Scaling Efficiency:}
        \begin{equation}
        E_w = \frac{T(n, P)}{T(n', 1)} \quad \text{where } n' = \frac{n}{P}
        \label{eq4}
        \end{equation}
        Weak scaling measures how well execution time is maintained when both problem size and processor count grow proportionally. A perfectly scalable algorithm would exhibit $E_w \approx 1$. However, in many parallel TSP solvers, weak scaling efficiency degrades beyond a certain processor count due to synchronization and load balancing issues \cite{yavits2014effect, bertuletti2023fast}.
\end{itemize}

% \paragraph{Scalability in Heuristic and Hybrid Methods.}

Unlike exact solvers, \textbf{heuristic} and \textbf{hybrid metaheuristic} approaches generally exhibit better scalability due to their flexible and loosely coupled architectures:

\begin{itemize}
    \item \textbf{Heuristic Solvers:}  
    Population-based heuristics scale well under both strong and weak scaling regimes. Independent fitness evaluations, solution updates, and decentralized data structures reduce inter-process communication. For example, in ACO, each ant builds a tour independently, allowing scalable execution across multiple cores or GPUs—though pheromone updates introduce some synchronization overhead.

    \item \textbf{Hybrid approaches} often retain scalability by parallelizing individual components. Fitness evaluation, local search, and global search strategies can be offloaded to separate threads or processors. However, coordination between modules (e.g., when combining exact local search with global heuristics) can create synchronization points that limit scalability. 
    
    % \hl{Recent GPU-based hybrid frameworks have reported improved weak scaling through asynchronous design and memory reuse strategies} \hl{[citation needed]}.
\end{itemize}

Overall, \textbf{heuristic} and \textbf{hybrid methods} are more amenable to scalable parallel implementation than \textbf{exact algorithms}, especially on modern multicore and many-core platforms. However, achieving high efficiency in large-scale settings still requires careful attention to load balancing, memory access patterns, and communication design.

\subsubsection{\textbf{Solution Optimality}}
\label{sec5.1.3}

Solution quality in TSP optimization is typically assessed by how closely the obtained solution approaches the known optimal value. For small and medium-sized instances where the optimal tour cost $Z^*$ is known (e.g., from Concorde), the solution quality can be quantified using the \textbf{optimality gap}:

\begin{equation}
\text{Optimality Gap (\%)} = \left( \frac{Z - Z^*}{Z^*} \right) \times 100\%
\label{eq5}
\end{equation}

Where $Z$ is the solution obtained by the algorithm, and $Z^*$ is the known optimal solution. For \textbf{exact algorithms}, this gap is zero by definition, as they guarantee finding the globally optimal tour. Studies such as \cite{Korte2008} confirm that exact algorithms maintain $100\%$ accuracy, but often at exponential computational cost.

In contrast, \textbf{heuristic methods} do not guarantee optimality and are instead evaluated based on how close their solutions are to $Z^*$. These approaches trade off solution quality for execution speed and scalability. Nevertheless, many heuristics are capable of producing near-optimal solutions for large instances with significantly lower computational requirements.

Some \textbf{Hybrid metaheuristics} have shown superior solution quality compared to standalone heuristics. For example, \cite{hu2007hybrid} demonstrated that combining evolutionary strategies with local improvement routines substantially reduces the optimality gap while maintaining reasonable runtime. This is particularly effective when the local search is applied selectively to high-quality solutions within a population.

For large-scale TSP instances where $Z^*$ is unknown, relative performance is often assessed using the best-known solution or averaged across multiple heuristic runs \cite{zhou2025dualopt,Rego2011}. In such cases, benchmarking against trusted solvers (e.g., Concorde for smaller instances or widely reported best solutions) provides a practical baseline for evaluating heuristic accuracy.

\subsubsection{\textbf{Memory Consumption}}
\label{sec5.1.4}

Memory consumption ($M$) is a critical constraint in solving the Traveling Salesman TSP, particularly for \textbf{exact algorithms} in parallel environments. It refers to the total memory required during execution and can significantly impact the feasibility and scalability of an algorithm.

\textbf{Exact algorithms}, especially those based on dynamic programming or Integer Linear Programming (ILP), are often memory-intensive. With dynamic programming, classical formulations such as the Held-Karp algorithm \cite{held1962dynamic} require storing the cost of visiting subsets of cities, resulting in exponential space complexity.

    \begin{equation}
    M(n) = \mathcal{O}(n^2 \cdot 2^n)
    \label{eq6}
    \end{equation}

    This renders the approach impractical for instances beyond a few dozen cities due to exponential memory growth. Also, with ILP-based solvers \cite{catanzaro2015improved}, memory usage grows with the number of variables and constraints, typically in the order of:

    \begin{equation}
    M(n) = \mathcal{O}(n^2)
    \label{eq7}
    \end{equation}

    While less memory-intensive than DP, ILP solvers must store large coefficient matrices, branching information, and intermediate LP relaxations, which can become a bottleneck on large instances.

In parallel implementations, memory is often distributed across multiple processors. This distribution alleviates some pressure on individual nodes but introduces additional overhead from inter-process communication and memory synchronization. As noted by \cite{burkhovetskiy2017parallelizing, linderoth2001parallel}, this communication overhead can significantly affect overall performance, particularly in distributed-memory systems. Efficient memory management and locality-aware design are essential to maintain scalability in large-scale parallel TSP solvers.

In parallel \textbf{exact algorithms}, memory is typically distributed across processors to reduce local memory pressure. However, this introduces communication overhead and memory synchronization issues that can degrade performance, especially in distributed-memory systems \cite{burkhovetskiy2017parallelizing, linderoth2001parallel}. Efficient memory layout, data locality, and communication minimization are essential for scalability.

In parallel \textbf{exact algorithms}, memory is typically distributed across processors to reduce local memory pressure. However, this introduces communication overhead and memory synchronization issues that can degrade performance, especially in distributed-memory systems \cite{burkhovetskiy2017parallelizing, linderoth2001parallel}. Efficient memory layout, data locality, and communication minimization are essential for scalability.

Unlike exact algorithms, \textbf{heuristic} and \textbf{hybrid metaheuristic} approaches are generally lightweight in terms of memory usage. They do not require storing combinatorially large state spaces or constraint matrices, making them suitable for large TSP instances.

On one hand, \textbf{heuristic algorithms} typically only store current candidate solutions, a population of routes (in GA), or pheromone matrices (in ACO). Their memory requirements are generally more $\mathcal{O}(n^2)$ or less, and scale linearly with the number of agents or population size \cite{chitty2018applying, alba2005parallel}. For example, ACO maintains a pheromone matrix of size $n \times n$, which is memory-efficient even for thousands of cities.

On the other hand, \textbf{hybrid methods} combine global search strategies (e.g., GA or ACO) with local improvement heuristics (e.g., 2-opt, Lin-Kernighan). While they introduce more intermediate data (e.g., local search buffers), they remain memory-efficient. Many implementations benefit from GPU parallelism by minimizing data movement and sharing local search results across threads \cite{abdelkafi2014parallel, melab2013gpu}. However, memory usage can increase if large populations or multiple local search modules are maintained in parallel \cite{skinderowicz2020implementing}.

Thus, \textbf{heuristic} and \textbf{hybrid} solvers offer a favorable trade-off: they achieve competitive solution quality with significantly lower memory requirements, making them more scalable for large TSP instances. This advantage becomes even more pronounced in parallel environments where memory per core is limited.

\subsubsection{\textbf{Parallel Efficiency ($PE$)}}
\label{sec5.1.5}

Parallel efficiency is a core performance metric that evaluates how effectively an algorithm utilizes available processors in a parallel computing environment. It is particularly relevant in the context of solving the TSP using \textbf{exact}, \textbf{heuristic}, and \textbf{hybrid metaheuristic} methods.

$PE$ is defined as the ratio of actual speedup to the number of processors, and is mathematically expressed as:

\begin{equation}
PE = \frac{S}{P} = \frac{T(n, 1)}{P \cdot T(n, P)}
\label{eq8}
\end{equation}

\noindent where $T(n, 1)$ is the execution time on a single processor, $T(n, P)$ is the execution time on $P$ processors, and $S$ is the speedup. Ideally, $PE = 1$ indicates perfect scaling—i.e., doubling the number of processors halves the execution time. In practice, however, $PE < 1$ due to factors such as synchronization overhead, communication latency, load imbalance, and memory contention.

% \paragraph{Parallel Efficiency Across TSP Solver Types.}
% \begin{itemize} \item 
\textbf{Exact Methods:} Although theoretically applicable, parallel efficiency is not a widely reported metric for exact solvers due to their inherent sequential components and data dependencies. For example, parallel branch-and-bound algorithms often exhibit sharp efficiency decline beyond 16 processors, primarily due to overhead in synchronizing shared bounds and distributing subproblems \cite{maher2021assessing}. This limits the scalability of exact solvers for large TSP instances.

\textbf{Heuristic algorithms} are more amenable to parallelization. Independent operations like fitness evaluation or ant path construction can be performed in parallel, enabling significant speedup. Studies show that well-implemented parallel heuristics can achieve up to 3$\times$ speedup on multi-core systems; however, efficiency declines as the complexity of the fitness function increases \cite{wei2021multi}.

\textbf{Hybrid approaches} are well-suited for GPU acceleration. CUDA-based GA-ACO hybrid implementations have been reported to reach over 80\% parallel efficiency, particularly when load balancing and memory access are optimized \cite{abdelkafi2014parallel}. These models often achieve a strong trade-off between computational efficiency and solution quality.

$PE$ is essential for evaluating the scalability and practicality of TSP solvers in high-performance computing environments. While \textbf{exact methods} face fundamental limits in scalability, \textbf{heuristic} and \textbf{hybrid methods}—especially those using GPUs or multicore architectures—achieve superior efficiency due to their modular and loosely coupled structures. As a result, GPU-accelerated hybrid metaheuristics represent one of the most promising directions for future research in large-scale TSP optimization.

\subsection{\textbf{Exact Algorithm Metrics}}
\label{sec5.1.6}

In parallel TSP optimization, exact algorithms such as Branch and Bound (B\&B), Integer Linear Programming (ILP), and Dynamic Programming (DP) are critical for obtaining provably optimal solutions. However, these algorithms incur high computational costs and memory usage, especially when scaling to larger instances. As a result, evaluating their feasibility in parallel settings requires targeted metrics that assess computational efficiency, scalability, and pruning effectiveness.

\paragraph{\textbf{Branching Factor (BF)}}
\label{sec5.1.6.1}

The branching factor is a key metric in evaluating the efficiency of B\&B and DP methods. It quantifies the average number of child nodes generated from each parent node during the search process. A higher branching factor implies a larger search tree, which increases both time complexity and memory overhead \cite{maher2021assessing}.

In B\&B, a high BF leads to an increased number of subproblems and reduced pruning efficiency. Efficient bounding strategies—such as dual bounds or problem-specific heuristics—can significantly reduce BF, thereby improving search performance \cite{zhang2000depth}. In DP, BF contributes to state-space explosion, particularly in formulations like the Held-Karp algorithm. Although ILP solvers do not utilize BF explicitly, they rely on constraint relaxations and cutting planes to manage solution space complexity \cite{zhang2000depth}.

\subsection{\textbf{Heuristic-Based Method Metrics}}
\label{sec5.1.7}

Heuristic algorithms provide near-optimal solutions with much lower computational overhead compared to exact methods. These solvers are particularly well-suited for large-scale parallel TSP applications due to their flexibility and scalability. To evaluate their performance, several metrics are used to capture robustness, convergence behavior, and diversity maintenance.

\subsubsection{\textbf{Heuristic Robustness Index (HRI)}}
\label{sec5.1.7.1}

The Heuristic Robustness Index (HRI) measures the consistency of a heuristic’s performance across diverse TSP instances. It reflects the algorithm’s stability in delivering high-quality solutions despite variations in problem size, topology, and constraints. HRI is defined as the normalized variance of solution quality across a benchmark suite:

\begin{equation}
HRI = \frac{1}{n} \sum_{i=1}^{n} \left( \frac{Q_i - \bar{Q}}{\bar{Q}} \right)^2
\label{eq9}
\end{equation}

where $Q_i$ is the solution quality for instance $i$, and $\bar{Q}$ is the average quality across all $n$ instances. Lower HRI values indicate higher robustness. For example, the Lin–Kernighan heuristic has been shown to perform reliably across a wide range of TSP instances \cite{rego2011traveling}.

\subsubsection{\textbf{Convergence Rate}}
\label{sec5.1.10.1}

Convergence rate measures the number of iterations required for a heuristic algorithm to stabilize its solution. Unlike exact methods, which explore all feasible solutions, heuristics converge toward good solutions through iterative improvement. Convergence can be modeled as:

\begin{equation}
C = \mathcal{O}(f(n, \theta))
\label{eq12}
\end{equation}

where $n$ is the problem size and $\theta$ represents algorithm-specific parameters such as population size, mutation rates, or temperature schedules. Adaptive heuristics—such as GAs with dynamic mutation rates—often exhibit faster and more stable convergence in large-scale TSP instances \cite{wang2008adaptive}.

\subsubsection{\textbf{Diversity Metrics: \textit{Entropy}, \textit{Spread Factor}, and \textit{LOAR}}}
\label{sec5.1.10.2}

Maintaining solution diversity is essential in heuristic methods to avoid premature convergence and local optima traps. Common diversity metrics include entropy, spread factor, and the Local Optima Avoidance Rate (LOAR).

\paragraph{\textit{\textbf{Entropy}}}

Entropy quantifies the unpredictability or diversity in a population and is commonly measured using the Shannon–Wiener Diversity Index:

\begin{equation}
E = - \sum_{i=1}^{R} p_i \log(p_i)
\label{eq13}
\end{equation}

Here, $p_i$ denotes the proportion of individuals in the $i$-th category, and $R$ is the total number of categories. Higher entropy indicates a more diverse population, which improves exploration and reduces the chance of stagnation \cite{dat2024hsevo}.

\paragraph{\textit{\textbf{Spread Factor}}}

The spread factor evaluates how widely solutions are distributed in the search space. It helps assess an algorithm's exploratory capacity:

\begin{equation}
SF = \frac{1}{N} \sum_{i=1}^{N} \sqrt{(x_i^j - \bar{x}^j)^2}
\label{eq14}
\end{equation}

Where $N$ is the population size, $x_i^j$ is the $j$-th coordinate of individual $i$, and $\bar{x}^j$ is the population centroid in that dimension. A high spread factor suggests a more even distribution of candidate solutions, which can delay convergence but improve global search \cite{kshirsagar2016hybrid, e2012new}.

\paragraph{\textit{\textbf{LOAR}}}

Though less formally defined, LOAR captures how frequently a heuristic algorithm avoids getting trapped in local optima. It is often estimated based on repeated runs and the proportion of unique or globally competitive solutions discovered. High LOAR is correlated with both high entropy and spread factor, especially in rugged search landscapes.

\subsection{\textbf{Hybrid Metaheuristics Metrics}}
\label{sec5.1.8}

Hybrid metaheuristics combine multiple optimization strategies—often from different algorithmic paradigms—to leverage their complementary strengths. Unlike standalone exact or heuristic methods, hybrid approaches are evaluated based on their ability to integrate components effectively, adapt to varying problem complexities, and exploit parallel architectures. This section outlines two important metrics for assessing hybrid metaheuristics in parallel TSP solvers.

\subsubsection{\textbf{Hybrid Integration Ratio (HIR)}}
\label{sec5.1.8.1}

One of the main challenges in designing hybrid metaheuristics is determining how much each component contributes to the overall solution quality. The HIR quantifies this contribution by comparing the hybrid model’s performance to that of its best-performing standalone component. It is defined as:

\begin{equation}
HIR = \left( \frac{Q_{\text{hybrid}} - Q_{\text{best-standalone}}}{Q_{\text{best-standalone}}} \right) \times 100
\label{eq10}
\end{equation}

Where $Q_{\text{hybrid}}$ is the solution quality achieved by the hybrid algorithm, and $Q_{\text{best-standalone}}$ is the highest solution quality among the individual components (e.g., GA, ACO, or SA). A higher HIR value indicates that the hybrid model effectively enhances performance beyond the capabilities of its components.

For instance, \cite{rokbani2021bi} reported a GA–ACO hybrid achieving 99\% accuracy, surpassing both GA and ACO individually. Similarly, \cite{saxena2017parallelizing} demonstrated that CUDA-based hybrid models achieved speedups exceeding 80\%, highlighting the impact of parallel architectures on hybrid model efficiency.

\subsubsection{\textbf{Algorithmic Synergy Index (ASI)}}
\label{sec5.1.8.2}

The algorithmic synergy index evaluates how effectively the integrated algorithms in a hybrid model collaborate to improve performance beyond the sum of their parts. It compares the performance of the hybrid solver to the sum of individual performances in isolation:

\begin{equation}
ASI = \frac{P_{\text{hybrid}}}{\sum_{i=1}^{n} P_{\text{individual}_i}}
\label{eq11}
\end{equation}

Where $P_{\text{hybrid}}$ is the performance of the hybrid algorithm (e.g., solution quality or fitness), and $P_{\text{individual}_i}$ is the performance of the $i$-th standalone component. An ASI value:

\begin{itemize}
    \item $> 1$ suggests a \textit{synergistic effect}, where the hybrid outperforms the additive benefits of its components.
    \item $< 1$ indicates a \textit{redundant or ineffective combination}, where the hybrid performs worse than the sum of its parts.
\end{itemize}

This metric is particularly useful in evaluating cooperative designs, such as those inspired by heterosis theory or biologically motivated co-evolution. For example, \cite{cai2024cooperative, senkerik2020brief} proposed cooperative hybrid metaheuristics that exploit algorithmic heterogeneity to enhance global optimization performance, directly aligning with the concept of synergy evaluation.

Together, HIR and ASI provide a comprehensive lens for evaluating hybrid metaheuristics in parallel TSP optimization. While HIR emphasizes performance improvement over standalone baselines, ASI evaluates the emergent behavior from algorithmic integration. These metrics help researchers identify both successful hybridization strategies and areas where component coordination may require refinement.

% \subsubsection{Other metrics}
% \label{sec5.1.10}
% \textcolor{red}{[update text below]} 
\section{Open Challenges and Future Directions}
\label{sec_gap}

Despite substantial progress in parallel optimization for the Traveling Salesman Problem (TSP), numerous open challenges persist that continue to shape the trajectory of research in this field. A foundational difficulty lies in achieving \textbf{effective load balancing} within parallel implementations. In heterogeneous environments, uneven distribution of computational tasks leads to suboptimal resource utilization and reduced performance. Compounding this are persistent \textbf{memory bandwidth constraints}, especially prevalent in exact methods like dynamic programming and integer linear programming, which demand intensive memory access and often hit scalability ceilings. Optimizing memory layout and reducing latency are thus critical for handling large-scale instances efficiently. Additionally, the \textbf{generalizability of heuristics} remains a central concern—many algorithms require intricate, instance-specific tuning. The need for adaptive, self-configuring strategies capable of generalizing across diverse TSP instances is increasingly urgent for practical scalability.

Parallel \textbf{exact methods} offer theoretical guarantees of optimality but face severe limitations in real-world scalability. Empirical evaluations have shown that these methods often suffer from \textbf{memory saturation}, even in distributed environments, as the problem's state space grows exponentially \cite{burkhovetskiy2017parallelizing, linderoth2001parallel}. Further, \textbf{scalability deteriorates} with increasing problem size due to synchronization overheads and communication costs among processing units \cite{al2023parallel}. Despite their precision, the \textbf{exponential time complexity} of exact algorithms renders them computationally infeasible for very large instances \cite{Korte2008}. Consequently, their application is typically restricted to small-to-moderate TSP instances, reinforcing the need for more scalable heuristic or hybrid solutions.

In contrast, parallel \textbf{heuristic methods} demonstrate significant scalability and speed advantages, particularly when executed on GPUs. Genetic Algorithms (GAs), for instance, exhibit strong acceleration patterns under GPU-based parallelization. However, not all heuristics benefit equally—algorithms like Ant Colony Optimization (ACO) often face \textbf{diminishing returns} due to global dependencies such as pheromone table synchronization. Moreover, \textbf{parameter sensitivity} is a recurring obstacle, as heuristic performance is frequently influenced by the choice and tuning of hyperparameters. While self-adaptive mechanisms offer a promising mitigation strategy \cite{rego2011traveling}, these methods can still encounter \textbf{scalability limits}, including memory bottlenecks and coordination overheads, in large-scale distributed environments.

To bridge these gaps, parallel \textbf{hybrid metaheuristics} have emerged as a powerful paradigm for large-scale combinatorial optimization. By leveraging distributed and GPU-based parallelism, hybrid models reduce computational burden and improve solution quality \cite{hu2007hybrid}. Empirical studies have shown GPU-accelerated hybrids achieving speedups exceeding 40× in ACO implementations and runtime reductions of over 80\% in CUDA-based systems \cite{ismail2024gpu, gohil2022travelling}. These architectures enhance robustness and mitigate premature convergence through diversification and redundancy \cite{almeida2025revisiting}. Models that combine GAs with ACO, or use hybrid OpenMP-MPI-CUDA frameworks, have demonstrated superior scalability and solution fidelity on TSP instances with thousands of nodes \cite{rokbani2021bi, dahiya2018literature}.

The integration of machine learning (ML) into these hybrid metaheuristics introduces novel capabilities but also distinct challenges. One key issue is the \textbf{generalization gap}, where models trained on small or narrow datasets fail to scale or adapt to unseen instances \cite{joshi2019efficient}. Additionally, \textbf{inference time} remains a limiting factor, as complex deep learning architectures are often too slow for real-time applications. Training-related concerns, such as \textbf{slow convergence rates}, further hinder deployment. Finally, the success of \textbf{ML-heuristic hybridization} depends on whether such integrations genuinely improve performance beyond what either method can achieve alone. Addressing these limitations is essential to realize ML’s potential in high-performance optimization.

Recent advancements signal promising \textbf{emerging research directions}. Graph Neural Networks (GNNs), for example, have demonstrated effectiveness in learning structural properties of graphs and enhancing local search convergence when combined with heuristics \cite{hudsongraph, joshi2019efficient}. Reinforcement learning, when integrated with classical solvers as in reinforced hybrid GAs, has shown superior performance in large-scale TSPs \cite{ouyang2024generalization}. Meanwhile, multi-objective optimization frameworks—such as Graph Pointer Networks trained under reinforcement learning—have enabled solvers to address trade-offs between distance, time, and other constraints \cite{Alexandria2024}.

Nevertheless, unresolved issues remain. These include the \textbf{lack of standardized evaluation metrics} for ML-based solvers, which impedes cross-study comparisons; \textbf{scalability bottlenecks} due to load imbalance, memory contention, and communication latency; and the \textbf{limited adaptability} of deep models to varying instance characteristics. Future work should focus on \textbf{integrating deep learning with metaheuristics} for dynamic, instance-aware parameter tuning; \textbf{developing quantum-classical hybrid algorithms} that combine quantum exploration with classical global search; and \textbf{designing robust benchmarking frameworks} with diverse datasets and evaluation tools tailored for AI-augmented and quantum-enabled solvers.

In conclusion, the interplay between exact, heuristic, and hybrid metaheuristic approaches continues to define the frontier of TSP optimization. Addressing the technical, algorithmic, and architectural challenges outlined above will be critical in developing next-generation solvers that are not only faster and more scalable but also more adaptive, intelligent, and capable of operating across diverse real-world scenarios.

\section{Conclusion}
\label{sec_conc}

The TSP remains one of the most studied combinatorial optimization problems due to its computational complexity and broad applicability in logistics, transportation, and network routing. This paper presented a comparative review of parallel TSP optimization techniques, with the aim of evaluating the efficiency, scalability, and practical applicability of exact, heuristic, hybrid, and machine learning-based approaches in parallel environments.

The review shows that parallel computing has significantly enhanced the capability of TSP solvers to address large-scale instances. \textbf{Exact algorithms}, while offering guaranteed optimality, are hindered by exponential time and memory requirements, limiting their utility to small and mid-sized problems. In contrast, \textbf{heuristic methods}—especially those accelerated using GPUs—offer considerable speed improvements while maintaining acceptable solution quality. Among these, \textbf{hybrid metaheuristics} demonstrate superior performance by effectively combining global search and local refinement strategies. Emerging machine learning-based solvers also show promise but currently face limitations in generalization and scalability.

These findings suggest that hybrid and hardware-accelerated methods offer the most practical balance between efficiency and accuracy for large-scale TSP optimization. The insights gained from this review highlight the importance of architectural alignment, adaptive algorithm design, and benchmark standardization in advancing parallel optimization research.

Future work should focus on developing adaptive and self-tuning hybrid frameworks that integrate deep learning components for improved generalization, as well as exploring quantum-inspired and quantum-classical hybrid models for solving increasingly complex TSP instances. Additionally, the creation of standardized benchmarks and evaluation metrics for AI-augmented and parallel solvers will be essential for ensuring reproducibility and fair comparison across methodologies.

% In summary, parallelism has become a cornerstone in TSP research, and its synergy with emerging computing paradigms presents exciting opportunities for building scala

%% references
\bibliographystyle{elsarticle-num}  
\bibliography{references}

\end{document}